\documentstyle[epsfig,preprint,eqsecnum,aps]{revtex}

\begin{document}
\setcounter{section}{1}
\draft

\author{A. G. C. Haubrich, D. A. Wharam, H. Kriegelstein, S. Manus, A. 
Lorke,\\ and J. P. Kotthaus} 
\address{Sektion Physik der LMU, Geschwister-Scholl-Platz 1, D-80539 
M\"unchen, Germany} 
\author{A. C. Gossard} 
\address{Materials Department, University of California, Santa Barbara, 
California 93106}

\title{Parallel Quantum-Point-Contacts as High-Frequency-Mixers} 
\date{\today}
\maketitle

\begin{abstract}

  We present the results of high-frequency mixing experiments 
  performed upon parallel quantum point-contacts defined in the 
  two-dimensional electron gas of an 
  Al$_{x}$Ga$_{1-x}$As/GaAs-hetero\-structure.  The parallel geometry, 
  fabricated using a novel double-resist technology, enables the 
  point-contact device to be impedance matched over a wide frequency 
  range and, in addition, increases the power levels of the mixing 
  signal while simultaneously reducing the parasitic source-drain 
  capacitance.  Here, we consider two parallel quantum point-contact 
  devices with 155 and 110 point-contacts respectively; both devices 
  operated successfully at liquid helium and liquid nitrogen 
  temperatures with a minimal conversion loss of 13 dB.

\end{abstract}



Since the discovery of the quantized conductance in ballistic quantum 
point-contacts (QPCs) \cite{wees,david} these devices have been the 
subject of a large number of theoretical and experimental 
investigations.  Recently, a number of papers have focussed upon the 
high-frequency (HF) device aspects of such devices.  For example, the 
suggestion that photon-assisted tunneling (PAT) may be observed in 
QPCs \cite{hu,fedi} has led to a number of experimental investigations 
\cite{wyss,janssen} although conclusive proof of this phenomenon in a 
single QPC remains elusive.  In this Letter we concentrate on the 
microwave mixing properties of an integrated parallel QPC device in
the ballistic regime and demonstrate that it operates as a sensitive and
efficient mixer.

At low temperatures the elastic mean free path in a high-mobility two 
dimensional electron gas (2DEG) can be as large as $10 \mu m$ and the 
electronic transport through a mesoscopic device such as a QPC 
therefore ballistic.  Consequently the associated transit time through 
the active region of the device can be of the order of $1 ps$. It has 
therefore been suggested \cite{kelly2} that the high-frequency 
operating limit of such devices could be as high as $10 THz$.  
Furthermore, it has been predicted \cite{kelly} that a QPC should 
exhibit pronounced non-linearities at small source-drain bias due to 
the saturation of the current carrying states within the QPC. 
Experimentally the observed non-linearity is less pronounced due both 
to the self-consistent modification of the subband occupancy 
\cite{tommy} as well as to the thermal broadening at the Fermi energy.  
Nevertheless, we have previously demonstrated \cite{goedel} at 
frequencies up to $10 GHz$ that a single QPC can be operated 
successfully as a sensitive high-frequency mixer.

The main problem associated with using QPCs as HF-mixing devices lies 
in the magnitude of the device impedance at the operating point.  
Typically these devices operate best when only a few one-dimensional 
subbands are occupied and consequently the real part of the impedance 
is of the order of several $k\Omega$.  For the single QPC device 
previously investigated \cite{goedel} the impedance matching of the 
device to the external $50 \Omega$ experimental setup was achieved 
using an impedance transformation with discrete, lithographically 
defined components.  The disadvantage of this method of impedance 
matching lies in its narrow bandwidth which, in the above experiment, 
limited the device performance.  An alternative approach to the 
impedance matching of the device is to be found in the parallel 
integration of a finite number of QPCs.  The resistive component of 
the device impedance clearly scales inversely with the number of QPCs 
and broadband matching over a wide range of operating conditions is 
readily obtained.  This solution offers the additional advantage of 
increased power levels in the mixing signals and also the automatic 
reduction of the parasitic source-drain capacitance.  For the parallel 
structures considered in this Letter we compare the performance of two 
devices with 155 and 110 parallel QPCs.

For the fabrication of a parallel QPC device, the conventional lateral 
geometry, based upon the split-gate technique \cite{thornton}, is not 
suitable.  Instead we have designed a single gate structure as 
illustrated schematically in fig.\ \ref{fig:modgate} where the 
distance of the Schottky gate from the semiconductor surface is 
periodically modulated by thin, insulating strips of negative resist.  
Under the influence of a negative gate-bias this distance modulation 
translates into an effective electrostatic modulation along the length 
of the gate, and hence to the formation of parallel QPCs with a 
lithographic width and length of $200 nm$.  In the heterostructure 
material used for these experiments the 2DEG is situated at a depth of 
$46 nm$; low-temperature transport measurements for this material
yield an elastic scattering time and length of $6.1 ps$ and $1.6 \mu m$
respectively. It is found that an applied bias of $-200 mV$ is sufficient 
to define the parallel QPCs and that their width, and hence 
resistance, can be continuously tuned to pinch-off which occurs at 
approximately $720 mV$.  Prior experiments on bulk gate devices have 
shown that the negative resist material used acts efficiently as an 
insulator and that depletion of the 2DEG occurs first for applied 
biases beyond $-2 V$; hence we conclude that the depletion of the 
conducting channel within the individual QPCs occurs primarily via 
lateral depletion, and that the transport through the QPCs is ballistic.

The mesoscopic sample was included in an optimized microstrip geometry 
with impedance matched regions for the source and drain contacts, 
which permitted the simultaneous measurement of two parallel QPC 
devices.  The gate coupling was achieved via an air-bridge technique 
\cite{hur} designed to isolate the Schottky electrode from the HF 
signal.  The completed sample was then mounted in an external 
microstrip setup connected to semi-rigid coaxial cables and inserted 
into the bath of a low-temperature cryostat.  The impedance matching 
of the experimental setup was investigated using voltage standing wave 
ratio measurements as a function of both the microwave frequency as 
well as the voltage applied to the gate electrode.  Both parallel QPC 
devices showed the same qualitative and quantitative bahaviour; the 
impedance increased with the applied gate bias from values close to 
$50 \Omega$ at definition to roughly $150 \Omega$ at pinch-off.  
Furthermore a systematic increase in impedance with frequency was 
observed which we attribute to the parasitic, non-resistive elements 
of our setup.  We conclude that we have achieved reasonable impedance 
matching over the range of operating conditions relevant for these 
devices.

The quantitative evaluation of the performance of both devices was 
achieved while operating the QPCs as reflection mixers in the small 
signal limit.  Two HF generators, operating at $2.9 GHz$ and $2.45 
GHz$ respectively, were coupled using a power-divider and the HF 
signal applied to the source contacts via a bias-T and directional 
coupler.  The reflected signal was measured at the intermediate 
frequency $450 MHz$ using a spectrum-analyser.  The signal power level 
was held constant at $-35 dBm$ for all experiments while the power 
level of the local oscillator was varied from $-35 dBm$ up to $-2 
dBm$.  Typical experimental results are shown in fig.\ 
\ref{fig:lcvsd155} where the power level of the intermediate signal is 
plotted as a function of the applied source-drain bias for a fixed 
gate-voltage of $-0.5 V$.  The observed mixing signals have clear 
maxima at absolute values of source-drain bias around $65 mV$.  The 
simultaneously measured current-voltage characteristics are shown in 
fig.  \ref{fig:kenn} for typical gate-voltages used in the 
measurements.  According to the simple saturation theory \cite{kelly} 
the maximum non-linearity should occur around $E_{f}/e$ which in our 
sample corresponds to $10 meV$.  At small local-oscillator levels we 
observe a second maximum in the mixing signal at roughly $V_{SD} = 
10mV$, however at higher power levels the broad maximum dominates the 
mixing characteristic.  This dominant signal is associated with the 
self-consistent source-drain characteristics\cite{tommy}, as is the 
asymmetry of the observed signal.  The optimal conversion loss was 
determined experimentally by varying the local oscillator power and a 
minimum value of $C = 13 dB$ attained at $L_{0} = -2 dBm$.  This value 
is in reasonable agreement with a small-signal analysis of the 
simultaneously recorded current-voltage characteristic.

The device performance is summarised in fig.\ \ref{fig:cont155} where 
the intermediate signal power is plotted as a function of the 
gate-voltage and source-drain bias for fixed local oscillator and 
signal power levels of $-20 dBm$ and $-35 dBm$ respectively.  The 
broad maximum at a gate-voltage of $-0.55 V$ shows that the device is 
operating best when the individual QPCs are well defined.  A naive 
estimate of the channel width of each QPC at this gate-voltage 
suggests that roughly one or two subbands are contributing to the 
current.  Beyond pinch-off, which occurs at $V_{G}=-0.72 V$ and is 
clearly visible in the data, the effectiveness of the device is 
drastically impaired.  This clearly shows that the mixing properties 
are related to the QPC channels themselves and are not a manifestation 
of a FET structure.

The results for the parallel QPC device with 110 QPCs are both 
qualitatively and quantitatively very similar; interestingly the 
intermediate signal level was consistently $2 dB$ larger than the 155 
QPC device.  This effect was due to the better impedance matching at 
the operating point, and more work is required to determine the 
optimum number of parallel QPCs.  As has been previously noted 
\cite{goedel} the mixing properties of QPCs result essentially from 
the width modification of the active channel and are not directly 
related to the subband structure of the QPC itself.  As such, it is to 
be expected that the parallel QPC devices may be operated at higher 
temperatures when the subband structure is thermally broadened.  This 
expectation has been confirmed by measurements performed at $77 K$ 
upon the two devices; both devices operated successfully with 
conversion losses only $2 dB$ larger than at $4 K$.

In conclusion, we have fabricated parallel QPC devices using  a novel 
double-resist technique, and have shown that they can be 
operated as high-frequency mixers with reasonable conversion loss for 
the configuration considered.  It remains to be seen whether these 
devices could be operated successfully with the HF-signal applied to 
the gate electrode. Ideally, this geometry would better exploit the 
transistor action of the QPC device, however the impedance matching 
of the gate presents a significant technological problem.

We acknowledge financial support from the Deutsche 
Forschungsgemeinschaft.  The LMU-UCSB collaboration is supported by a 
Max-Planck research award.



%

\begin{figure}
\begin{center}
\leavevmode 
\bigskip
\caption{The parallel 
device geometry used in these experiments is schematically 
illustrated.  The Schottky gate height is periodically modulated via 
the insulating resist strips defined on the surface of the 
semiconductor using high-resolution electron-beam lithography.  }
\label{fig:modgate}
\end{center}
\end{figure}

\begin{figure}
\begin{center}
\leavevmode 
\caption{The 
intermediate frequency power level is plotted as a function of the 
applied source-drain bias for different local oscillator power 
levels ($-35 dBm$ to $-2 dBm$ -- corresponding to the lowest and the 
highest trace) at fixed gate-voltage.  The signal 
power level is fixed at $-35 dBm$, the temperature of the sample is 
$4.2 K$.}
\label{fig:lcvsd155}
\end{center}
\end{figure}

\begin{figure}
\begin{center}
\leavevmode 
\caption{Current-voltage-characteristics for different gate-voltages 
under microwave bias 
are shown for the relevant voltage-region from -0.3V up to -0.9V.}
\label{fig:kenn}
\end{center}
\end{figure}

\begin{figure}
\begin{center}
\leavevmode 
\caption{A summary of the conversion-loss data is shown in a contour-plot 
as a function of the applied source-drain and gate voltages. The 
results shown are for the device with 155 QPCs.}
\label{fig:cont155}
\end{center}
\end{figure}

\newpage

\begin{figure}
\begin{center}
\leavevmode 
\epsfig{file=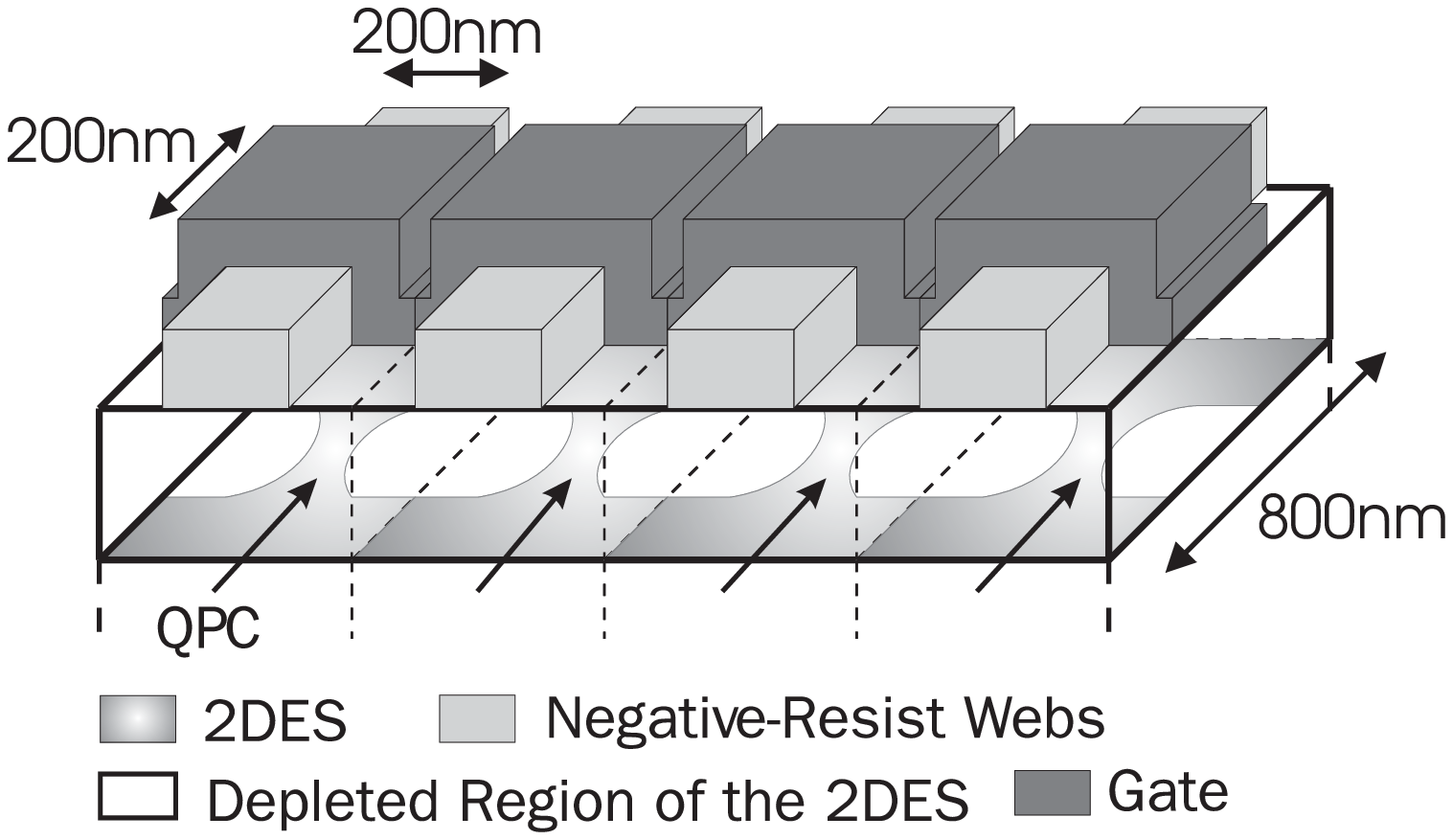,width=14cm,keepaspectratio} \bigskip
\end{center}
\end{figure}
\vspace{11.5cm}
\noindent
A.G.C Haubrich et al. \hspace{2cm} fig. 1 \\
Applied Physics Letters

\newpage

\begin{figure}
\begin{center}
\leavevmode 
\epsfig{file=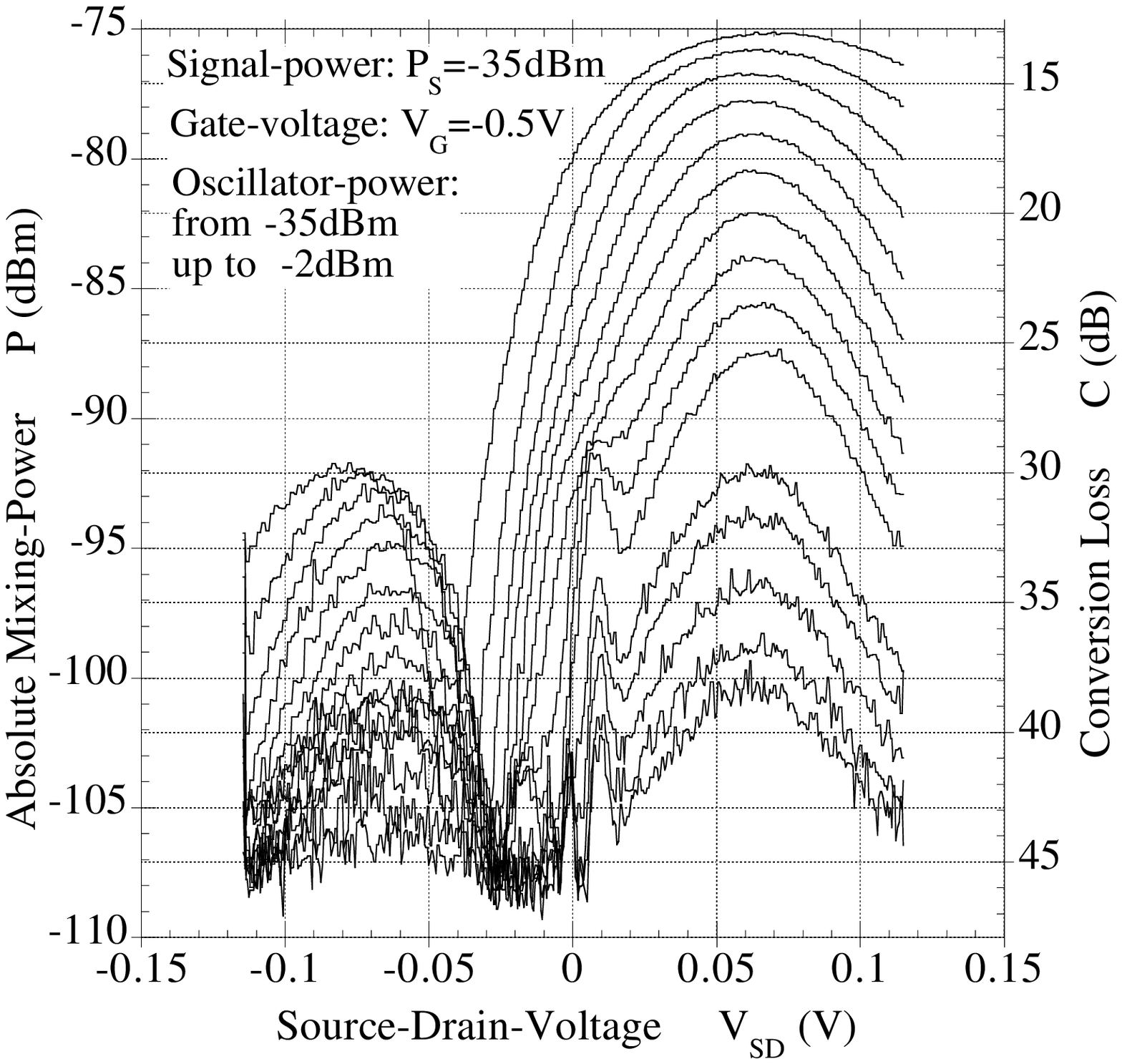,width=14cm,keepaspectratio} 
\end{center}
\end{figure}
\vspace{6cm}
\noindent
A.G.C Haubrich et al. \hspace{2cm} fig. 2 \\
Applied Physics Letters

\newpage

\begin{figure}
\begin{center}
\leavevmode 
\epsfig{file=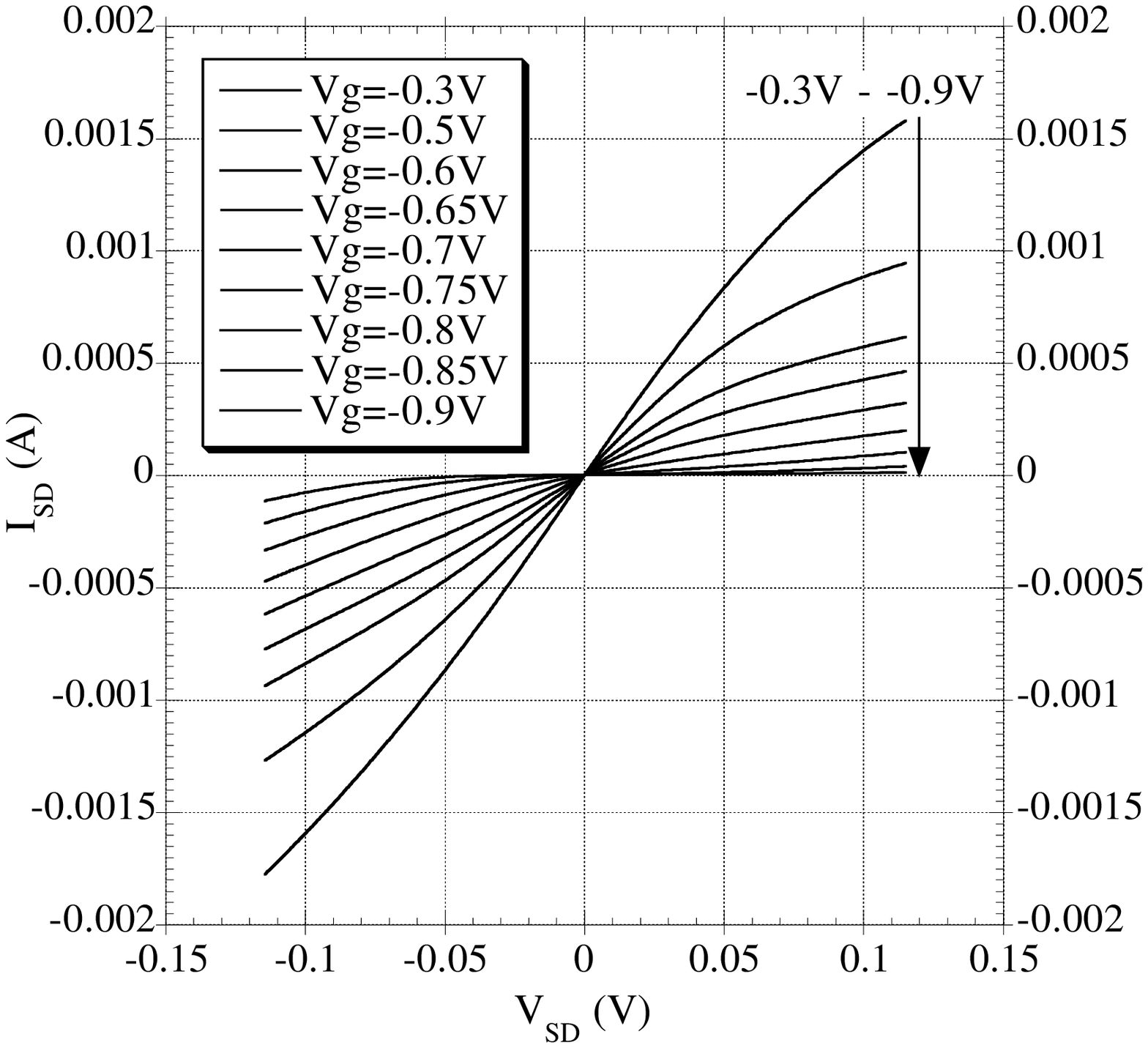,width=14cm,keepaspectratio} 
\end{center}
\end{figure}
\vspace{6cm}
\noindent
A.G.C Haubrich et al. \hspace{2cm} fig. 3 \\
Applied Physics Letters

\newpage

\begin{figure}
\begin{center}
\leavevmode 
\epsfig{file=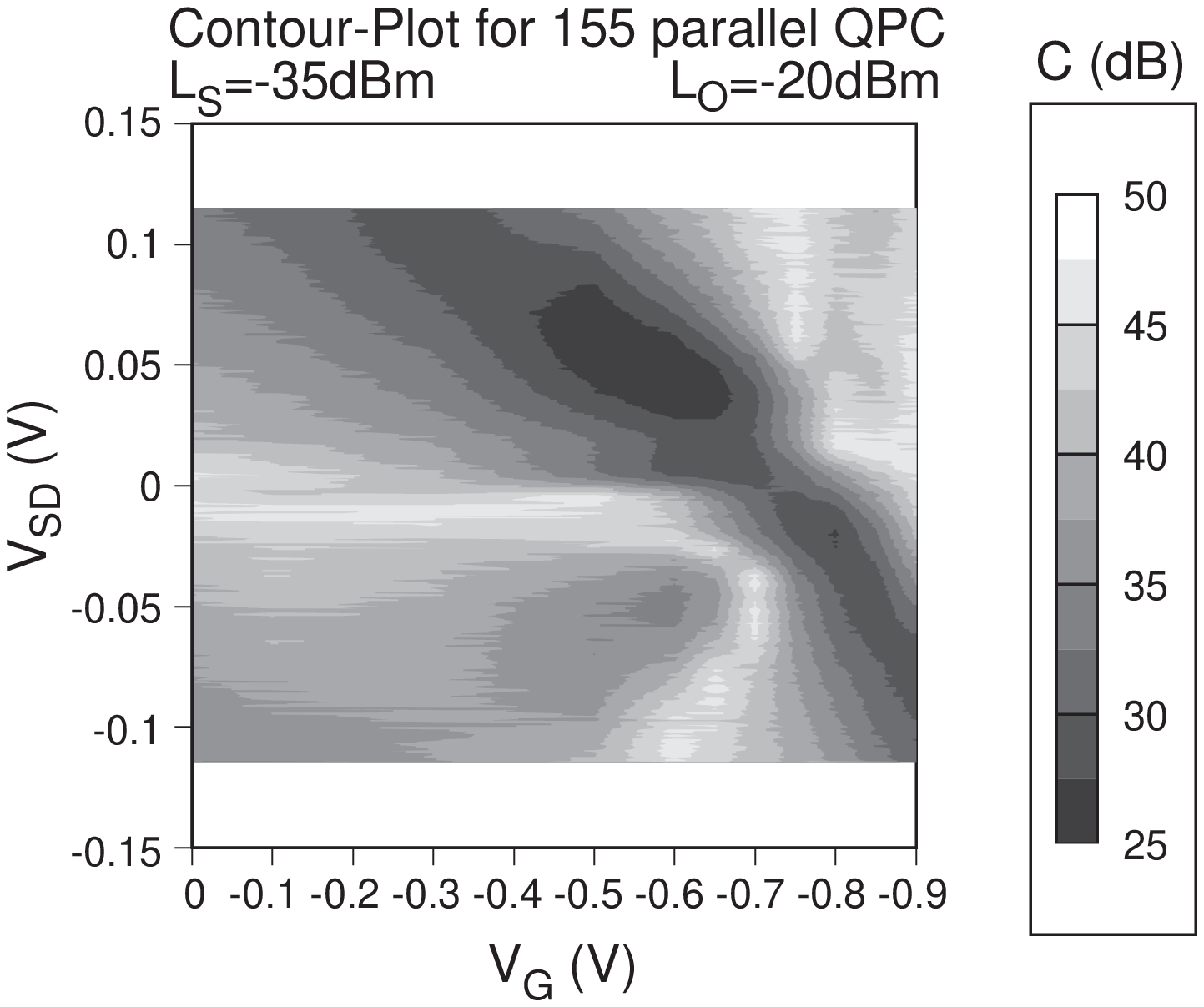,width=14cm,keepaspectratio} 
\end{center}
\end{figure}
\vspace{8.5cm}
\noindent
A.G.C Haubrich et al. \hspace{2cm} fig. 4 \\
Applied Physics Letters


%
%

\end{document}